\documentclass[twocolumn,amsmath,amssymb,aps,pre]{revtex4}

\usepackage{graphicx}
\usepackage{dcolumn}
\usepackage{bm}

\newcommand{\alat}{a_l}
\newcommand{\beq}{\begin{equation}}
\newcommand{\gcc}{\mbox{g cm$^{-3}$}}
\newcommand{\eeq}{\end{equation}}
\newcommand{\req}[1]{Eq.~(\ref{#1})}
\newcommand{\mion}{m_\mathrm{ion}}
\newcommand{\nion}{n}

\newcommand{\aB}{a_B}
\newcommand{\kD}{\kappa_{D}}
\newcommand{\kTF}{\kappa_\mathrm{TF}}
\newcommand{\pF}{p_F}
\newcommand{\vF}{v_F}
\newcommand{\EF}{E_F}
\newcommand{\UJ}{U_J}
\newcommand{\UL}{U_L}
\newcommand{\UM}{U_M}

\begin{document}

\title{Electrostatic energy of Coulomb crystals 
with polarized electron background}\thanks{Published in \textit{Phys.\
Rev. E} \textbf{103}, 043205 (2021), \\
\url{https://doi.org/10.1103/PhysRevE.103.043205}}

\author{A. A. Kozhberov}
\email{kozhberov@gmail.com}

\author{A. Y. Potekhin}
\email{palex@astro.ioffe.ru}

\affiliation{Ioffe Institute, Politekhnicheskaya 26, Saint Petersburg, 194021, Russia}%

\date{\today}

\begin{abstract}

Outer crusts of neutron stars and interiors of cool white dwarfs consist
of bare atomic nuclei, arranged in a crystal lattice and immersed in a
Fermi gas of degenerate electrons. We
study electrostatic properties of such Coulomb crystals, taking into
account the polarizability of the electron gas and considering 
different structures, which can form the ground state: body-centered
cubic (bcc), face-centered cubic (fcc), hexagonal close-packed (hcp), and
MgB$_2$-like lattices. At zero temperature the electrostatic energy
brings a fundamental contribution to the total energy of the classical
Coulomb crystal, which allows us to study structural transitions in the
neutron-star crusts and crystallized white-dwarf interiors. To take the
electron background polarization into account, we use the linear
response theory with the electron dielectric function given either by
the Thomas-Fermi approximation or by the random-phase approximation
(RPA). We compare the widely used nonrelativistic (Lindhard) version of
the RPA approximation with the more general, relativistic (Jancovici)
version. The results of the different approximations are compared to
assess the importance of going beyond the Thomas-Fermi or Lindhard
approximations. We also include contribution of zero-point vibrations of
ions into the ground-state energy. We show that the bcc lattice forms
the ground state for any charge number $Z$ of the atomic nuclei at the
densities where the electrons are relativistic ($\rho\gtrsim10^6$ \gcc),
while at the nonrelativistic densities ($\rho \lesssim 10^6$ \gcc) the
fcc and hcp lattices can form the ground state. The MgB$_2$-like lattice
never forms the ground state at realistic densities in the crystallized
regions of degenerate stars. The RPA corrections strongly affect the
boundaries between the phases. As a result, transitions between
different ground-state structures depend on $Z$ in a nontrivial way. The
relativistic and quantum corrections produce less dramatic effects,
moderately shifting the phase boundaries.

\begin{description}
\item[PACS numbers]
52.27.Gr, 52.27.Lw, 52.25.Kn, 05.70.Ce
\end{description}

\end{abstract}

\maketitle

\section{Introduction}

A model of Coulomb
plasma assumes that the system consists of
point-like charged particles and the
neutralizing background. 
The basic physics behind this model was
explored in many works and summarized in a number of review papers,
textbooks and monographs (e.g., \cite{YSh88,Fortov16,EbelingFF17},
and references therein).
This model is quite universal and is used in
many areas of physics, for example, in astrophysics, for description of
such different objects as neutron stars, white dwarfs, gas giants, and
dusty clouds (e.g., \cite{BH54,ST83,Kit,FKM04,Fort1,HPY07,A10}).

At sufficiently low temperatures or high densities, the Coulomb plasma
forms a crystal (see, e.g., Ref.~\cite{Bonitz_08}
for a discussion of the phase diagram
of Coulomb crystals in classical and quantum plasmas).
In particular, some regions of the degenerate stars,
namely crusts of neutron stars (e.g., \cite{HPY07}, and references therein)
and cores of sufficiently
cool white dwarfs (e.g., \cite{HPY07,vH19}, and references therein), 
are formed by Coulomb crystals, where a 
lattice of bare atomic
nuclei is embedded in the background of degenerate electrons.
The question, which type of crystal lattice forms in such stars, is
not solved completely. For crystals with the same ion charge number $Z$
it was discussed in \cite{B01,HPY07,CF16,K20}. More complex systems with
two or three different charged ions were studied in
\cite{KB12,CF16,K18,K20}. In all of the papers cited above, the uniform
electron background was considered. 

The aim of the present paper is to take into account the influence of
the electron background polarization on the electrostatic energy and the
ground state of a Coulomb crystal.  Previously this problem was studied
only for the body-centered cubic (bcc) and face-centered cubic (fcc)
lattices \cite{PH73,PC00,YSh88,B02,K182}. In this paper, we
investigate other lattices, such as hexagonal close-packed (hcp) and
one-component MgB$_2$. It allows us to construct a more complete picture
of structural transitions in degenerate stars. As well as the previous
authors, we assume that the electron screening is weak, which is
justified in the case of strongly degenerate electrons. 
We consider different lattices, calculate their
ground-state energies, and find which lattice is most preferable. The
electron polarization effects are described in frames of the linear
response theory by the dielectric function, for which we use and compare
three approximations: the widely used Thomas-Fermi theory, the
Lindhard model \cite{Lindhard} based on the nonrelativistic random-phase approximation
(RPA), and the extension of the Lindhard
model for the relativistic electrons \cite{J62} (see also, e.g., 
\cite{AshcroftMermin,YSh88}). Comparing the results
obtained using these three models, we will estimate the importance of
the corrections beyond the Thomas-Fermi approximation and of the effects of
special relativity.

The paper is organized as follows. In Sec.~\ref{sec:param} we introduce
basic parameters of the polarized electron background and present the
three different approximations for the dielectric function, which are
used in the subsequent sections. In Sec.~\ref{sec:elstat} we determine
the ground state structure of the Coulomb plasma, following the approach
of Chamel and Fantina \cite{CF16}. According to this approach, at zero
temperature the type of the lattice is mainly determined by the
electrostatic energy, all point-like ions being in equilibrium positions
$\bm{R}_l$. We compute the electrostatic energy in different
approximations and compare the results for the phase boundaries between
different ground-state structures. In Sec.~\ref{sec:zero-point} we go
beyond the approach of Chamel and Fantina \cite{CF16} by taking into
account the zero-point vibrations of ions around their equilibrium
positions. In Sec.~\ref{sec:other} we check the accuracy of the
calculated ground-state energy and positions of phase boundaries by
evaluation of possible corrections to our approximations. Results are
discussed and conclusions are given in Sec.~\ref{sec:concl}.

\section{Parameters and approximations for polarized background}
\label{sec:param}

Since ions move slowly, it is sufficient to use the linear response
approximation and the static longitudinal dielectric function
$\epsilon(q)$ to describe the degenerate electron background
\cite{HS71}. In the approximation of the uniform (``rigid'') background,
$\epsilon(q)=1$. In the
present paper we take the background nonuniformity into account
by considering only the first-order perturbation corrections to
the dielectric function $\epsilon(q)$. It is convenient to write
\begin{equation}
\epsilon(q)=1+\frac{\kTF^2}{q^2}\epsilon_2(q)~,
\label{eps}
\end{equation}
where $q$ is the wavenumber, $\epsilon_2(q)$ is the correction,
which will be discussed later, and 
\begin{equation}
\kTF = \sqrt{4\pi e^2 \frac{\partial n_e}{\partial\mu_e}}
\approx
  2\sqrt{\frac{e^2}{\pi\hbar \vF}}
     \,\frac{\pF}{\hbar}
\label{kTFdegen}
\end{equation}
is the Thomas-Fermi electron wavenumber.
Here, 
$n_e$ is the electron number density,
$\mu_e$ is the electron chemical potential,
 $\pF=\hbar(3\pi^2 n_e)^{1/3}$ is the Fermi momentum,
and $\vF=\partial \EF/\partial \pF$ is the Fermi velocity,
$\EF$ being the Fermi energy. The first equality in \req{kTFdegen}
is quite general, while the second is approximately valid in
the case of strongly degenerate electrons, where $\mu_e \approx \EF$.

In the degenerate stars, the electrons can be relativistic 
due to high densities.
The relativity parameter \cite{Salpeter61} is
\begin{equation}
x \equiv \frac{\pF}{m_e c} \approx 0.01\,(\rho\, Z/A)^{1/3},
\label{xrel}
\end{equation}
where $m_e$ is the electron mass, $\rho$ is mass density in units of
\gcc, $Z$ is the ion charge number and $A$ is the relative atomic weight
of the considered isotope. The second part of \req{xrel} assumes the
neutrality condition $n_e = Z \nion$, where $\nion$ is the ion number
density.

In plasma physics, it is customary to use the dimensionless density
parameter $r_s = a_e/\aB$, where 
$a_e = (4\pi n_e/3)^{-1/3}$ is the electron sphere radius
and $\aB\equiv \hbar^2/m_e e^2$ is the Bohr radius. 
It is related to the relativity parameter as
$r_s=(9\pi/4)^{1/3}\alpha/x\approx0.014/x$, where $\alpha\equiv e^2/\hbar c$ is the fine structure constant.

In general, \req{kTFdegen} is valid both for the
the nonrelativistic treatment and for the special relativity taken into
account. In the latter case, \req{kTFdegen} 
leads to the expression
\beq
   \kTF\,a_e = \left(\frac{18}{\sqrt{\pi}}\right)^{\!\!1/3}
    \sqrt{\frac{\alpha\gamma}{x}}
   \approx 0.185\sqrt{\frac{\gamma}{x}},
\label{kTFae}
\eeq
where $\gamma\equiv \sqrt{1+x^2}$ is the electron Lorentz factor on the Fermi
surface.  

The
importance of the electron polarization is mainly indicated by the
Thomas-Fermi parameter $ \kTF a$, where $a\equiv (4\pi \nion/3)^{-1/3}$
is the ion sphere radius.  The charge neutrality requires that
$a=a_e Z^{1/3}$, and then \req{kTFae} yields
\beq
\kTF a \approx0.185\,Z^{1/3}\sqrt{\frac{\gamma}{x}}.
\label{kTFa}
\eeq

The model of uniform electron background is equivalent to neglecting the
term with $\epsilon_2(q)$ in \req{eps}. To go beyond this approximation,
we use three expressions
for $\epsilon_2(q)$. The most common of them is the Thomas-Fermi
approximation, which is valid in the long-distance (short wavenumber) limit at any $x$.
It gives
\begin{equation}
\epsilon_\mathrm{2TF}(q)=1.
\label{epTF}
\end{equation}

In the non-relativistic case ($x \ll 1$) and for strongly degenerate
electrons, the RPA leads to the
Lindhard (L) dielectric function  \cite{Lindhard},
\begin{equation}
\epsilon_\mathrm{2L}(q)=
\frac{1}{2}+\frac{1-y^2}{4y}\ln\left|\frac{1+y}{1-y}\right|~. 
\label{epL} 
\end{equation}
where $y=\hbar q/(2 \pF)\approx0.26qa_e$.

The RPA with allowance for the electron relativity leads to the
Jancovici (J) model \cite{J62}
\begin{eqnarray}&&\hspace*{-1em}
\epsilon_\mathrm{2J}(q) =
 \frac{2}{3}-\frac{2}{3}\frac{y^2x}{\gamma}\ln(x+\gamma)
+
\!\frac{x^2+1-3x^2 y^2}{6yx^2} \ln\!\left|\frac{1+y}{1-y}\right|
\nonumber\\&&\qquad
+
 \frac{2y^2 x^2-1}{6y x^2}\frac{\sqrt{1+x^2 y^2}}{\gamma}
      \ln \left|
      \frac{y\gamma+\sqrt{1+x^2 y^2}}{y\gamma-\sqrt{1+x^2 y^2}}
      \right|.
 \label{epJ}
\end{eqnarray}
The Jancovici dielectric function reduces to the Lindhard dielectric
function in the nonrelativistic limit $x\to0$, and both Lindhard and
Jancovici models reduce to the Thomas-Fermi model in the long-wavelength limit
$y\to0$.

\section{Electrostatic energy}
\label{sec:elstat}

\subsection{Uniform background}

First let us outline the main results for the Coulomb crystals with the
uniform background. The electrostatic energy of such crystals can be
written as
\begin{equation}
\UM=N\frac{Z^2e^{2}}{a}\zeta~,
\label{Um}
\end{equation}
where $\zeta$ is called a Madelung constant, and $N$ is the total number
of ions. Accurate values of the Madelung constant for four lattices with the
lowest $\UM$ are quoted in Table~\ref{tab:coef} from the
previous works (see the recent review \cite{K20} and references
therein). 

Here we use the traditional geometry of the hcp lattice (e.g.,
\cite{NagaiFukuyama83}) with the distance between hexagonal layers
$h_0/2=\sqrt{2/3}\, \alat \approx 0.816497 \alat$, where $\alat$ is the
lattice constant and $h_0$ is the height of the primitive Bravais cell,
while for the one-component lattice of the MgB$_2$ type the
lowest-energy configuration is used. The electrostatic energy of the
Coulomb hcp lattice is the lowest for the lattice with $h_0/2$ replaced by
$h_{\min}/2 \approx 0.81782 \alat$, but this difference affects only the
7th significant digit of $\UM$ and is unimportant, as we show in
Sec.~\ref{sec:other} below. 

From Table~\ref{tab:coef} we see that the bcc lattice has the lowest
electrostatic energy. So its formation among all lattices with one type
of ion in the elementary cell and uniform background is most likely. For
the multi-component crystals, formation of other lattices is also
possible (e.g., \cite{CF16,K20}, and references therein).

\begin{table}
\centering
\caption{Madelung constant $\zeta$ in \req{Um} and coefficient
$\eta_\mathrm{TF}$ of the quadratic
approximation (\ref{U2TF})
for the bcc, fcc, hcp, and MgB$_2$ lattice types.
}
\label{tab:coef}
\begin{tabular}{lcc}
\hline 
lattice  & $\zeta$     & $\eta_\mathrm{TF} $ \rule{0pt}{2.5ex}\\
\hline
bcc      & $-0.895929255682$ & $-0.103732333707$ \rule{0pt}{2.5ex} \\
fcc      & $-0.895873615195$ & $-0.103795687531$ \\
hcp      & $-0.895838120459$ & $-0.103809851801$ \\
MgB$_2$  & $-0.89450562823~$ & $-0.104080616256$ \\
\hline
\end{tabular}
\end{table}

\subsection{Polarized background. Basic notations}

Expressions for the electron screening corrections to the
thermodynamic functions of the one-component plasmas have been derived
in frames of the thermodynamic perturbation theory by Galam and Hansen
\cite{GalamHansen}. In the particular case of
the one-component Coulomb crystal in the harmonic approximation,
the polarization correction to the electrostatic energy
can be written as \cite{B02}
\begin{eqnarray}
\Delta U_\mathrm{pol} &=&
 N \frac{Z^2 e^2}{a} \frac{3}{2N_\mathrm{cell}^2}
 \sideset{}{'}\sum\limits_{m} \frac{1}{\left(G_m a\right)^2}
\left[\frac{1}{\epsilon(G_m)}-1\right] \nonumber \\
&&\times
 \sum\limits_{p,p'}
e^{i\bm{G}_m(\boldsymbol{\chi}_p-\boldsymbol{\chi}_{p'})}~,
   \label{U} 
\label{Upol}
\end{eqnarray}
where $N_\mathrm{cell}$ is the number of ions in the elementary cell and
the sums run over all basis vectors $\boldsymbol{\chi}_{p}$ in the
elementary cell ($p$ and $p'$) and all reciprocal lattice vectors
$\bm{G}_m$ with the exception of $G=0$. The total electrostatic energy
with the polarization correction becomes
\begin{equation}
U = \UM + \Delta U_\mathrm{pol}.
  \label{Uall}
\end{equation}
Three approximations for the dielectric function, given by
Eqs.~(\ref{epTF}), (\ref{epL}), and (\ref{epJ}), yield three
different values of electrostatic energy, which we denote
$U_\mathrm{TF}$, $\UL$ and $\UJ$, respectively.

Equation~(\ref{eps}) is written in the linear response approximation, 
which is valid if the typical energy of electron-ion interaction
$Ze^2/a$ is small compared to the Fermi energy $\EF$. Since
\beq
   \frac{Ze^2/a}{\EF} =
      \frac{x^2}{\gamma(\gamma-1)}\, \frac{(\kTF a)^2}{3^{4/3} }
       < 0.463 \,(\kTF a)^2,
\eeq
the condition $Ze^2/a\ll\EF$ is satisfied to a good accuracy, provided
that $\kTF a \lesssim 1$. This condition is usually well fulfilled in
the crystallized regions of the degenerate stars. It justifies that
$\epsilon(q)$ in \req{eps} contains only polarization corrections
proportional to $\kTF^2$. Higher order corrections to the dielectric
function are unknown.  To estimate a possible effect of the higher-order
terms,  we expand the electrostatic energy in a series in powers of
$\kTF a$ up to the quadratic terms. Then \req{Upol} is reduced to
\begin{eqnarray}
U_\mathrm{2pol} &=& - N \frac{Z^2 e^2}{a} (\kTF a)^2 
 \nonumber  \\
&\times&
  \frac{3}{2N_\mathrm{cell}^2}  \sideset{}{'}\sum\limits_{m}
\frac{\epsilon_2(G_m)}{(G_m a)^4} \sum\limits_{p,p'}
 e^{i\bm{G}_m(\boldsymbol{\chi}_p-\boldsymbol{\chi}_{p'})}~.
\label{U2}
\end{eqnarray}
Energies $U_2 \equiv \UM + U_\mathrm{2pol}$ are denoted further in the text as
$U_\mathrm{2J}$, $U_\mathrm{2L}$ and $U_\mathrm{2TF}$ for the Thomas-Fermi,
Lindhard, and Jancovici models, respectively.

\subsection{Thomas-Fermi model}
\label{sec:TF}

The Thomas-Fermi model is the simplest and most widely used
approximation for describing the effects of electron polarization. It
was applied to studies of the Coulomb crystals in degenerate stars in
many works (e.g., \cite{PH73,PC00,B02} and references therein). However,
all previous investigations concerned only the bcc and fcc lattices.
Here we extend the consideration to two other lattice types, hcp and 
MgB$_2$.

As can be seen from Eqs.~(\ref{eps}), (\ref{epTF}), (\ref{Um}), and
(\ref{U}), for each lattice type the dimensionless ratio
$U_\mathrm{TF}/(N Z^2 e^2 a^{-1})$ depends only on the single parameter
$\kappa_\mathrm{TF}a$. 

To obtain the electrostatic energy, as can be seen from \req{U2},
one has to calculate slowly converging
sums like $\sum_m (-1)^m G_m^{-4}$.
In the Thomas-Fermi model, this difficulty can be overcome
by the Ewald rearrangement \cite{Ewald}, which reduces 
Eqs.~(\ref{Um}) and (\ref{U}) to 
\begin{eqnarray}&&\hspace*{-1em}
U_\mathrm{TF} =
NZ^2e^{2}
\bigg\{\frac{1}{N_\mathrm{cell}}\sum\limits_{l,p,p'}
\left(1-\delta_{\bm{R}_l0}\delta_{pp'}\right) \frac{E_{-}+E_{+}}{4Y_l}
\nonumber \\&&\qquad
-
\frac{\kTF}{2}\,\mathrm{erf}
\left(\frac{\kTF}{2\mathcal{A}}\right)
-\frac{\mathcal{A}}{\sqrt{\pi}}\,
\exp\left({-\frac{\kTF^2}{4\mathcal{A}^2}}\right)
-\frac{2\pi n}{\kTF^2}
 \nonumber \\&&\qquad
+
\frac{1}{N_\mathrm{cell}^2}\sum\limits_{m,p,p'}
 \frac{2\pi n}{G_m^2+\kTF^2}
\nonumber \\&&\qquad\quad\times
  \mathrm{exp}\left[-\frac{G_m^2+\kTF^2}{4\mathcal{A}^2}
  -i\bm{G}_m(\boldsymbol{\chi}_p-\boldsymbol{\chi}_{p'})
  \right]\bigg\},
\qquad
\label{UTF}
\end{eqnarray}
where $E_{\pm}={e}^{\pm\kTF Y_l}\,\mathrm{erfc}
\left(AY_l\pm {\kTF}/(2\mathcal{A})\right)$,
$\bm{Y}_l=\bm{R}_l+\boldsymbol{\chi}_p-\boldsymbol{\chi}_{p'}$,
$\mathrm{erf}(x)$ is the error function,
$\mathrm{erfc}(x)\equiv1-\mathrm{erf}(x)$, and $\mathcal{A}$ is an arbitrary
constant; a good
numerical convergence of both sums is provided
by $\mathcal{A}\approx 2/a$ \cite{B02}.

\begin{figure}
\center{\includegraphics[width=\columnwidth]{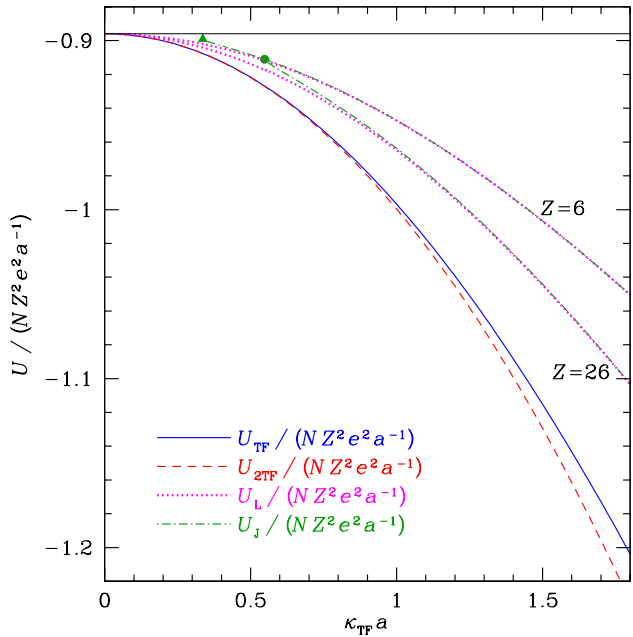}}
\caption{Electrostatic energy  of the bcc lattice in units of
$NZ^2e^2/a$ in different approximations. The full Thomas-Fermi
approximation $U_\mathrm{TF}$ (\req{UTF}, solid curve) and in the
second-order approximation $U_\mathrm{2TF}$ (\req{U2TF}, dashed curve)
are compared with the rigid background approximation $\UM$ (\req{Um},
the horizontal line) and with the more accurate approximations due to
Lindhard ($\UL$, Sec.~\ref{sec:L}, dotted lines) and Jancovici ($\UJ$,
Sec.~\ref{sec:J}, dot-dashed lines) for carbon ($Z=6$) and iron
($Z=26$). The triangle and the heavy dot mark the minimum of $\kTF a$ in
the relativistic (Jancovici) theory for C and Fe, respectively. }
\label{fig:1}
\end{figure}

In Fig.~\ref{fig:1}, the electrostatic energy $U$ is plotted by in
different approximations for the bcc lattice. In the chosen scale, the
difference between the results for different lattice types is not
noticeable. In the full Thomas-Fermi model, \req{UTF}, the bcc lattice has the lowest 
electrostatic energy at $\kTF a < 1.065714$, while the fcc lattice
has the lowest $U_\mathrm{TF}$ at larger $\kTF a$.

The situation changes if we take the first two terms of the series
expansion in the parameter $\kTF a$ only. This approximation gives
\begin{equation}
U_\mathrm{2TF}=N\frac{Z^2e^2}{a}(\zeta+\eta_\mathrm{TF}
 (\kTF a)^2)~, 
\label{U2TF}
\end{equation}
where constant $\eta_\mathrm{TF}$ depends on the type of the lattice and
is presented in the last column of Table~\ref{tab:coef}. At the selected
scale  in Fig.~\ref{fig:1} the difference between $U_\mathrm{TF}$ (the
solid line) and $U_\mathrm{2TF}$ (the dashed line) is noticeable only at
$\kTF a \gtrsim 1$. The difference between energies of different
lattices is very small, and the structural transition between the bcc
and fcc lattice moves from $\kTF a = 1.065714$ to  $\kTF a = 0.93715$,
furthermore at  $\kTF a > 1.58301$ the hcp lattice will possess the
smallest $U_\mathrm{2TF}$, while at $\kTF a >2.21838$ the MgB$_2$
lattice becomes more energetically preferable. However, with the
exception of the bcc-fcc transition, all other transitions are outside
the limits of the theory applicability.

The dotted and dot-dashed curves in Fig.~\ref{fig:1} show the electrostatic energy in more advanced approximations, which are discussed below.

\subsection{Lindhard model}
\label{sec:L}

The Lindhard expression for the dielectric function \cite{Lindhard} and
its generalization by Mermin \cite{Mermin} are widely used in the plasma
physics (e.g., \cite{GalamHansen,DornheimGB18}). The underlying RPA
approximation makes this model more accurate than the Thomas-Fermi
model.

In this case, the Ewald resummation cannot be performed explicitly.
Therefore we calculate the total electrostatic energy
$\UL$ according to Eqs.~(\ref{epL}), (\ref{U}),
and (\ref{Uall}). Unlike the Thomas-Fermi model, the single parameter
$\kappa_\mathrm{TF}a$ is not sufficient anymore. Since
$\epsilon_\mathrm{2L}(q)$ in \req{epL} contains an argument $y\propto q
a_e$, the second dimensionless parameter appears, $a_e/a$, which equals
$Z^{-1/3}$ due to the charge neutrality. In Fig.~\ref{fig:1}, the dotted
lines show $\UL$ for carbon and iron, which are most typical
chemical elements in the white dwarf cores and neutron star envelopes.
In both cases the polarization correction appears smaller than in the
Thomas-Fermi model. The difference $\UL - U_\mathrm{TF}$ is
smaller for the larger $Z$, which reflects the well known fact that the
Thomas-Fermi theory becomes more accurate with increasing $Z$.

Results for the ground-state structure are presented in
Fig.~\ref{fig:JL}, where the different regions hatched by the dashed
lines show the range of parameters $Z$ and $x$ at which the fcc or hcp
lattice has the lowest total energy, while the white area corresponds to
the bcc lattice. The MgB$_2$ lattice is never energetically preferred in
the density range shown in this figure, where formation of Coulomb
crystals can be expected in the degenerate stars. The right vertical
axis displays the density parameter $r_s$,  which is more customary than
$x$ in the nonrelativistic theory. The dot-dashed line corresponds to
$\kTF a=1.0657$, which describes the structural transition between the
bcc and fcc lattices in the Thomas-Fermi model.  Recalling that the
linear response theory is justified at $\kTF a \lesssim1$, we should
accept the results in the region below this line with caution. Anyway,
we see a remarkable difference between the results obtained using the
Thomas-Fermi and Lindhard models. For the latter model (unlike the
former one), the ground state structure depends on $Z$ in a nontrivial
way, if the density parameter $r_s\gtrsim0.01$ ($x\lesssim1.4$).
However, in any case the bcc lattice forms the true ground state at $r_s
< 0.01$.

\subsection{Jancovici model}
\label{sec:J}

The Lindhard expression for the dielectric function is derived in the
nonrelativistic formalism, therefore it is applicable only at $x\ll1$.
The Jancovici model \cite{J62} generalizes the Lindhard model taking the
effects of special relativity into account. Since the relativity
parameter $x$ is not small in the crystallized regions of the most
typical neutron stars and white dwarfs, the Jancovici model is more
suitable in the theory of these stars (e.g., \cite{YSh88,HPY07}),
although it is used relatively rarely (e.g., \cite{B02,PB99,DG09}). For
studying the electrostatic energy it was employed only in
Ref.~\cite{B02} for the bcc and fcc (but not hcp) lattices. 

\begin{figure}
\center{\includegraphics[width=.95\linewidth]{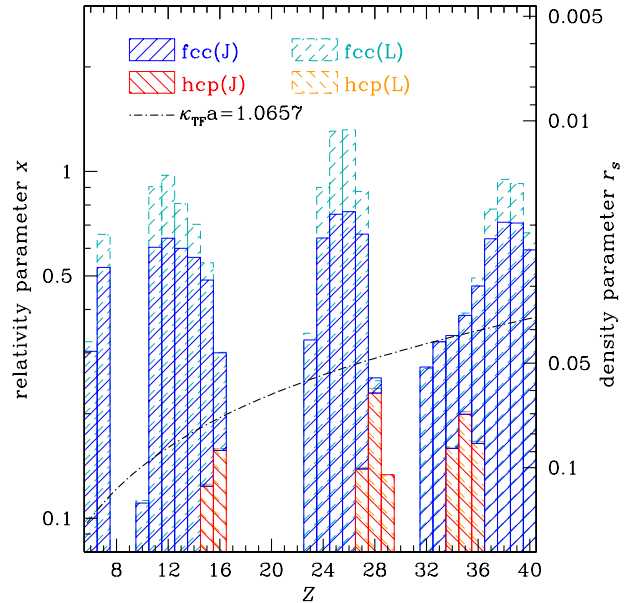}}
\caption{The ranges of charge number $Z$, relativity parameter $x$, and
density parameter $r_s$, at which the fcc or hcp lattice (hatched areas
according to the legend) form the ground state of a classical Coulomb
crystal with polarizable electron background. The dashed and solid
hatchings correspond to the Lindhard and Jancovici approximations to the
dielectric function of the electron background. In the white area, the
ground state is formed by the bcc lattice. The dot-dashed line
corresponds to $\kTF a = 1.065714$ and separates the regions where the
ground state is bcc (above the line, where $\kTF a$ is smaller) or fcc
(below the line, where $\kTF a$ is larger) 
in the Thomas-Fermi approximation.} 
\label{fig:JL}
\end{figure}

The electrostatic energy for the Coulomb plasma of carbon and iron,
calculated using the Jancovici model ($\UJ$), is shown in
Fig.~\ref{fig:1} by dot-dashed curves (see the data in \cite{suppl}).
These curves are not continued to $\kTF a=0$, because in the
relativistic theory the Thomas-Fermi  parameter cannot be smaller than
$(\kTF a)_{\min}=0.185\,Z^{1/3}$, according to \req{kTFa}. Hence the
electrostatic energy noticeably differs from the limit $\UM$ at any
density. We see that $\UL$ and $\UJ$ are almost identical at $\kTF a
\gtrsim 2(\kTF a)_{\min}$, but the results of the two models
substantially differ at smaller values of $\kTF a$; thus, the special
relativity effects reduce the polarization correction by $\sim30$\%  at
$\kTF a \sim (\kTF a)_{\min}$.

The areas hatched with the solid lines in Fig.~\ref{fig:JL} show the
ranges of parameters $Z$ and $x$ where the fcc or hcp lattices have the
lowest electrostatic energy according to the relativistic (Jancovici)
model of the dielectric function. They can be compared with the areas
hatched with the dashed lines, which show analogous regions according to
the nonrelativistic (Lindhard) model, which we considered in
Sec.~\ref{sec:L}. The difference between these two approximations is
noticeable only at $x \sim 1$. Qualitatively, the results are very
similar. It means that the relativistic corrections to the dielectric
function are not very important for the structural transitions of the
Coulomb crystals. It is quite expected since the structural transitions
occur mostly at $x\lesssim1$. Note that relativistic corrections are
rather insignificant also for the phonon spectra of Coulomb crystals
\cite{B02,K18}.

\begin{figure}
\center{\includegraphics[width=.95\linewidth]{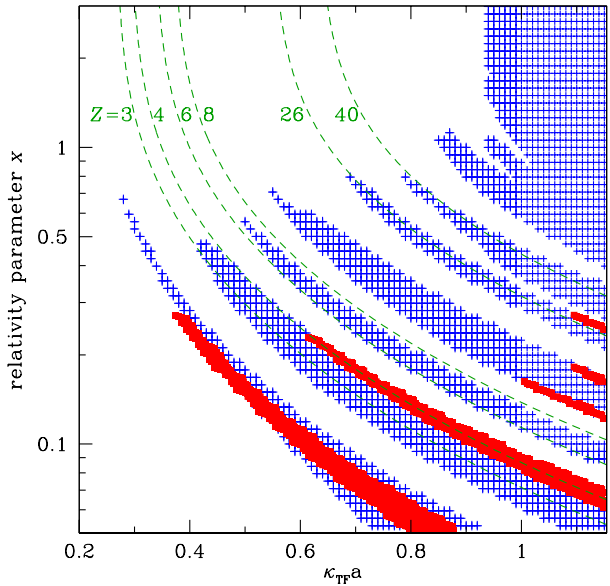}}
\caption{The ranges of parameters $x$ and $\kTF a$ at which one or
another lattice has the smallest $U_\mathrm{2J}$. The bcc lattice forms
the ground state in the white region, the fcc lattice in the blue
cross-hatched regions, and the hcp in the red filled regions.
The dashed lines corresponds to
$Z=3$, 4, 6, 8, 26, and 40 (marked near the curves).  
}
\label{fig:duj3}
\end{figure}

In Fig.~\ref{fig:duj3} we plot structural transitions for the
second-order approximation $U_\mathrm{2J}$. Here, $x$ and $\kTF a$ are
chosen as independent parameters. Our results for the bcc and fcc
lattices reproduce the results of Baiko \cite{B02}. At high $x$ the
transition between the bcc and fcc lattices takes place at $\kTF a$
between 0.935 and 0.945, which is consistent with the result of the
Thomas-Fermi approximation (it gives the transition at $\kTF a \approx
0.93715$).  At low densities the pattern of ground-state structures is
more complicated. In the latter case, there appear several domains in
the parameter plane where the hcp lattice is energetically preferable
according to the Jancovici model.

\section{Zero-point vibrations}
\label{sec:zero-point}

In the previous sections we did not consider zero-point quantum
vibrations of ions around their equilibrium positions. 
This model would be accurate for very massive ions. In
reality, the relative atomic weight $A$ varies from $A\approx2Z$ in the
outer shells of neutron star crusts or crystallized cores of white
dwarfs to $A\approx3.5Z$ near the neutron drip point in the neutron-star
crust. In this section, we consider the effects of the quantum
vibrations on the ground-state structure of a Coulomb crystal, assuming
$A=2Z$. In this way we obtain an estimate of the upper limits for these
effects, because a larger $A$ would result in a smaller vibration energy
and accordingly a smaller difference from the results obtained for the
classical (fixed in space) ions.

The energy of the ground-state quantum vibrations
of a three-dimensional harmonic lattice equals (cf., e.g., \cite{AshcroftMermin})
\beq
   E_0 = \frac{3}{2}N\hbar\langle\omega\rangle ,
\label{E0}
\eeq
where $\langle\omega\rangle$ is the phonon frequency
$\omega_\nu(\bm{k})$, averaged over all phonon modes $\nu$ and
wavevectors $\bm{k}$ in the first Brillouin zone.  In the first
approximation, the total ground-state energy is given by the sum $E = U
+ E_0$, where $U$ is the electrostatic energy (\ref{Uall}).
Equation~(\ref{E0}) can be written in the dimensionless form as
\beq
   \frac{E_0}{NZ^2e^2/a} =
    \frac{3}{2}\sqrt{\frac{3}{R_S}}\,u_1\,,
\label{E0a}
\eeq
where $R_S\equiv a/[\hbar/\mion(Z e)^2]\approx1823 A Z^{7/3}r_s$ is the
ion density parameter (analogous to $r_s$ for the electrons),
$u_1\equiv\langle\omega\rangle/\omega_p$,
$\omega_p= \sqrt{4\pi \nion Z^2 e^2 / \mion}$ is the ion
plasma frequency,
$\mion=Am_u$ is the ion mass, and $m_u$ is the unified
atomic mass unit.

To calculate $\langle\omega\rangle$ and thus $u_1$, the phonon spectrum
was obtained by solving the dispersion equation, as described in
Ref.~\cite{K18}: $\mathrm{det}\big\{D_{ss'}^{\alpha\beta}(\bm{k}) -
\omega_\nu^2(\bm{k})\delta^{\alpha\beta}\delta_{ss'} \big\} = 0$, where
$s$ and $s'$ run over the ions in the
elementary cell, and $D_{ss'}^{\alpha\beta}(\bm{k})$ is the dynamic matrix.
Generally, $D_{ss'}^{\alpha\beta}(\bm{k})$ is given, e.g., by Eq.~(5.6)
of Ref.~\cite{PH73}, but in this section we calculate it in the
Thomas-Fermi approximation (according to Eqs.~(3)\,--\,(5) of
Ref.~\cite{KB12}).

\begin{figure}
\center{\includegraphics[width=0.95\columnwidth]{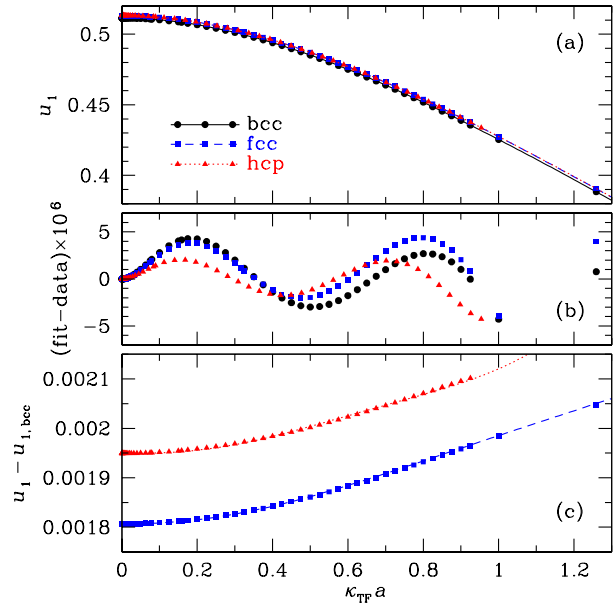}}
\caption{\textit{Top panel}: Dependence of the first phonon moment $u_1$
on the Thomas-Fermi parameter: computed values (black dots, blue
squares, and red triangles for the bcc, fcc, and hcp Coulomb lattices,
respectively) and the analytic fit, \req{u1fit} (the solid, dashed,
and dotted lines, respectively).
\textit{Middle panel}: The fit residuals, multiplied by $10^6$.
\textit{Bottom panel}: The excess of the first phonon moment $u_1$ for
the fcc and hcp Coulomb lattice over that for the bcc lattice,
calculated from the numerical results (the symbols) and the analytic fit
(the lines).
}
\label{fig:u1}
\end{figure}

The electron polarization decreases $u_1$ and slightly affects the
differences $\Delta u_1$ between the different lattice types. In the
Thomas-Fermi model this dependence is well described by the simple
Pad\'e approximation:
\beq
   u_1 = \frac{u_1^0\,\left[ 1+ p_4 (\kTF a)^3 \right] }{
      1+p_1 (\kTF a)^2 + p_2 (\kTF a)^4 + p_3 (\kTF a)^6}\,,
\label{u1fit}
\eeq
where $u_1^0$ is the value of $u_1$ in the one-component
plasma model with the rigid background and $p_i$ are fitting parameters,
which are given in Table~\ref{tab:u1}. The computed and fitted
dependencies of $u_1$ on $\kTF a$ are shown in Fig.~\ref{fig:u1}.  The
residuals between the fit, Eq. (\ref{u1fit}), and the numerical results
are smaller than $5\times10^{-6}$ (see the middle panel of
Fig.~\ref{fig:u1}), which provides a good analytic approximation to the
differences  $\Delta u_1$, which are typically a few times
$10^{-3}-10^{-4}$ in this approximation (the bottom panel). 

\begin{table}
\centering
\caption{Parameters of \req{u1fit}.
}
\label{tab:u1}
\begin{tabular}{lccccc}
\hline 
lattice& $u_1^0$ & $p_1$ & $p_2$ & $p_3$ & $p_4$ \rule{0pt}{2.5ex}\\
\hline
bcc    & 0.5113874 & 0.246236 & 0.05496 & 0.0026 & 0.08457 \rule{0pt}{2.5ex} \\
fcc    & 0.5131940 & 0.244996 & 0.05490 & 0.0025 & 0.08463 \\
hcp    & 0.5133369 & 0.245250 & 0.05663 & 0.0021 & 0.08599 \\
\hline
\end{tabular}
\end{table}

\begin{figure}
\center{\includegraphics[width=.95\linewidth]{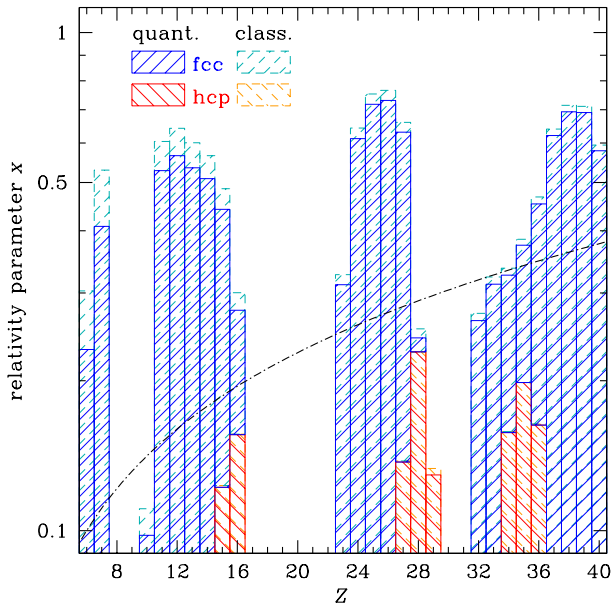}}
\caption{The same as in Fig.~\ref{fig:JL}, but taking the zero-point
ion vibrations into account (solid lines and hatched areas).
For comparison, the dashed lines and dashed hatched areas show the
results without the zero-point vibrations. The Jancovici model for the
electron dielectric function is used in all the cases.
} 
\label{fig:QJ}
\end{figure}

The zero-point energy given by Eqs.~(\ref{E0}) and (\ref{u1fit}) was
added to the electrostatic energy, computed according to Eqs.~(\ref{U})
and (\ref{Um}), to obtain the total ground-state energy $E$ of a Coulomb
crystal. Its absolute ground state is formed by the lattice that
delivers the lowest ground-state energy. The results of this evaluation
are shown in Fig.~\ref{fig:QJ}. Here we have assumed $A=2Z$. A larger
value of $A$ would lead to smaller differences from the classical-ion
model, whose results are also reproduced in this figure. We see that the
zero-point vibrations almost do not affect the boundaries of the hcp
domain. They somewhat decrease the densities of the fcc/bcc transition,
but this effect is not profound; it is is even smaller than the one due
to the special relativity corrections (cf.{} Fig.~\ref{fig:duj3}).

\section{Possible small corrections}
\label{sec:other}

One may note that the results presented in Figs.~\ref{fig:u1} and
\ref{fig:QJ} rely on the dependence of $u_1$ on the polarization
parameter $\kTF a$ obtained using the Thomas-Fermi model, which results
in the one-parameter dependence $u_1(\kTF a)$. A more accurate (e.g.,
Lindhard or Jancovici) model gives a two-parametric dependence $u_1(\kTF
a,x)$. An example of the bcc Coulomb lattice at $\kTF a=0.5$  presented
in Ref.~\cite{K183} indicates that the difference of the first phonon
moment in the Thomas-Fermi approximation $u_1$ relative to its 
rigid-background value $u_1^{(0)}$ may be slightly smaller in the
Jancovici model than in the Thomas-Fermi model. It is difficult to
evaluate the differences between $\Delta u_1$ between the different
lattice types beyond the Thomas-Fermi approximation with the accuracy
that is needed to determine the effect on the boundaries between
different ground-state lattice types, but the rough estimation for 
$\kTF a=0.5$ shows that $u_1$ of the fcc lattice is greater than  the
$u_1$ of the bcc lattice at any $x$. For instance, at $x=1$
$u_\mathrm{1,bcc} \approx 0.48587$  while $u_\mathrm{1,fcc} \approx
0.48686$.  Hence we expect that the $x$-dependence of $u_1$ cannot
qualitatively change the results presented in Fig.~\ref{fig:QJ}. To
check the last statement, we repeated our calculations in the extreme
case, where the polarization dependence of $u_1$ is completely
suppressed (i.e., $u_1=u_1^{(0)}$) and found that the boundaries between
different structural phases in Fig.~\ref{fig:QJ} shift no more than by
$\Delta\log_{10}x \lesssim0.0065$, provided that $x\gtrsim0.1$, which
means that the densities of phase boundaries vary by less than 1.5\%.
Clearly, such small shift of the boundary cannot have any observable
consequence, so that it is safe to use \req{E0} with the one-parametric
expression (\ref{u1fit}) for evaluation of the zero-point ion vibrations
energy.

As mentioned above, in the bulk of our numerical calculations we fixed
the height of the primitive cell in the hcp lattice at
$h_0=\sqrt{8/3}\,\alat$. However, the true minimum of the electrostatic
energy $U(h)$ as a function of the cell height $h$ is reached at a
slightly larger height value, $h_{\min}$. In the Thomas-Fermi
approximation for the polarizable electron background, $h_{\min}$ slowly
decreases with increasing $\kTF a$ and can be reproduced at $\kTF a
\lesssim3$ by the fitting expression 
\beq
   h_{\min} \approx
 h_0 + 0.00265\,\left\{1-\frac{0.155\,(\kTF a)^2
    }{1+0.11\,(\kTF a)^2}
    \right\}.
\eeq
The corresponding difference in the ground-state energy
is described by the fit
\beq
\Delta\zeta_h \equiv 
\frac{U(h_{\min}) - U(h_0)}{NZ^2e^2/a} \approx
 - 3.3\times10^{-7}\exp\bigg[-\frac{(\kTF a)^2}{2}\bigg].
\eeq
 At $\kTF a \lesssim3$ these expressions provide the absolute accuracy
within $\sim10^{-5}$ for $h_{\min}$ and within $\sim5\times10^{-9}$ for
$\Delta\zeta_h$. This correction shifts the densities of transitions
between different ground-state structural phases in Figs.~\ref{fig:JL}
and \ref{fig:QJ} by $\lesssim0.7$\%.

The MgB$_2$ lattice has a similar insignificant dependence $h_{\min}$ on
$\kTF a$ (see discussion in Ref.~\cite{K182}).

\section{Discussion and Conclusions}
\label{sec:concl}

We used and compared several theoretical models
to study the influence of the polarization of
degenerate electron background to the electrostatic properties of
Coulomb crystals and to their structure in the neutron star crusts
and white dwarf interiors.

The studies of the electrostatic energies of Coulomb crystals with
uniform background appear to be quite complete now, as presented in the
review \cite{K20}.  Based on the results of Ref.~\cite{K20}, we chose
four lattices with the lowest Madelung constant (bcc, fcc, hcp and
MgB$_2$) and investigated polarization effects for them. Since the
MgB$_2$ lattice never has the lowest energy the formation of other
lattices, which were not considered in the present work, is very
unlikely.

The Thomas-Fermi model predicts that at $\kTF a <1.065714$ the bcc
lattice forms the ground state, while at higher $\kTF a$ the fcc lattice
is energetically preferable. The Thomas-Fermi model, as well as the
linear response theory in general, allows one to obtain only corrections
$ \sim (\kTF a)^2$. Our estimate of the effect of the higher order
corrections shows that they can noticeably change the general picture of
structural transitions. For instance, omitting all higher order
corrections to the electrostatic energy within the Thomas-Fermi model
leads to the shift of the transition between bcc and fcc lattice to
$\kTF a \approx 0.93715$, while at $\kTF a \approx 1.58301$ transition
between the fcc and hcp lattices appear.

The Lindhard and Jancovici RPA-based models give quite similar results
as concerns the ground-state lattice structure, which qualitatively
differ from the Thomas-Fermi approximation. Instead of a single
structural transition from the bcc to fcc lattice, a strong nontrivial
dependence on the charge number $Z$ appears. In addition, it the hcp
lattice can form the ground state for some $Z$ at relatively low
density. The relativistic corrections to the dielectric function, given
by the Jancovici model, as well as the quantum corrections due to
zero-point vibrations of ions, only moderately shift the structural
transition densities. In all considered models, the bcc lattice
possesses the lowest electrostatic energy at the density parameters
$r_s<0.01$ (mass densities $\rho\gtrsim5\times10^6$ \gcc) for any $Z$.
In the most advanced of the models (Fig.~\ref{fig:QJ}), the
unconditional bcc stability range extends down to the relativity
parameters $x\gtrsim0.8$ (mass densities $\rho\gtrsim10^6$ \gcc).

As shown in Ref.~\cite{K182}, the equation for the electrostatic energy
of the Coulomb crystal with the Thomas-Fermi dielectric function is
almost the same as the equation for the electrostatic energy of the
Yukawa crystal, which is a usual model for ordered dusty plasmas
\cite{FH93,FH94I,FH94II,FHD97,FKM04,Fort1}. A slight difference is that
in the Yukawa crystal the screening parameter is the Debye wavenumber
$\kD$ and there is no restriction analogous to $\kTF a \lesssim 1$. The
Yukawa crystal model is used up to  $\kD a = 4.76$, when it becomes
unstable against phonon oscillations \cite{RKG88,K182}. Hence both
systems have the same structural transitions. Thus, while in the
Lindhard model at $\kTF a \gtrsim 1$ the fcc and hcp lattices become
energetically preferable and the strong dependence of structural
transitions on $Z$ appears, we can expect a similar situation if higher
order corrections to the Yukawa crystal model are taken into account.
Since in the dusty plasmas it is impossible to maintain strictly the
same charge for all grains, the formation of the hcp-fcc crystal mixture
is more likely.

The appearance of the hcp ground state at sufficiently large $\kTF a$ in
the linear response theory may be a key to explanation of data of the
experiment KPT-10 ``Kulonovskiy Kristall'' (``Coulomb Crystal'') onboard
the International Space Station. This experiment has revealed that dusty
particles with $\kD a = 0.5 - 3$ form an ordered system with the hcp and
fcc structures (e.g., \cite{Klumov_10,K10}). Since equations for the
electrostatic energy of the Coulomb crystal with the Thomas-Fermi
dielectric function and of the Yukawa crystal are the same, it makes
sense to assume that for systems with a non-degenerate background (such
as the dusty crystals) the higher order screening corrections may bring
us to the situation similar to the degenerate case, which is described
above, where the fcc and hcp lattices may form the ground state. 

It may be of interest for the reader to note that, beside the dusty
plasmas mentioned above, the normal electron-hole plasmas in
semiconductors also demonstrate some similarities to the dense
ion-electron plasmas. The crystallization of the electron-hole plasmas
with heavy holes was studied in Refs.~\cite{Filinov_07a,Filinov_07b},
using the path integral Monte Carlo simulations.

Back to degenerate stars, let us note that our reported model
improvements are likely unimportant for the ground-state equation of
state or chemical composition. Previously, Chamel and Fantina
\cite{CF16PRD} studied this problem based on the Thomas-Fermi model of
electron polarization, scaled so as to approach the more accurate
polarization results of Ref.~\cite{PC00}. In particular, Chamel and
Fantina found that corrections to the electron-capture threshold in
white-dwarf cores are very small and the neutron-drip density and
pressure in the crusts of neutron stars are only slightly shifted.
Although electron polarization may change the composition of the crust
of nonaccreting neutron stars, uncertainties in the masses of
neutron-rich isotopes were found to be more important than electron
exchange and polarization effects. As can be seen from our
Fig.~\ref{fig:1}, the polarization correction to the energy has the same
order of magnitude in the Thomas-Fermi model as in the more accurate
Lindhard and Jancovici models. Therefore, these improvements in the
model cannot qualitatively change the conclusions by Chamel and Fantina
\cite{CF16PRD}. Let us also note that some modern neutron-star equations
of state take sufficiently accurate account of the electron polarization
effects (e.g., Ref.~\cite{Pearson_18}).

On the other hand, our results show that the replacement of the
traditional Thomas-Fermi model by the RPA-based models of the
polarization corrections can strongly shift the boundaries between the
bcc and fcc lattice types and lead to appearance of the hcp lattices in
the cores of white dwarfs and crusts of neutron stars.  The interfaces
between different lattice structures may affect the kinetic properties
of the crust. For example, crust failure  due to accumulated stresses
during neutron-star evolution (e.g.,
Refs.~\cite{Ruderman91,UshomirskyCB00,LanderGourgouliatos19}) can
likely  develop along the surfaces separating the different lattice
types,  where the breaking stress can be smaller than its standard value
in the bulk of the crust \cite{KozhberovYakovlev20}. 

Based on the presented results, we can conclude that the most part of
the crust of neutron stars and crystallized cores of white dwarfs at
$\rho \gtrsim 10^6$ g/cm$^3$ consist of the bcc lattice at any $Z$,
which is the standard assumption for their modeling.  At lower
densities, however, the fcc and hcp lattices can form the true ground
state. Crystallization of the one-component plasma occurs at
$\Gamma\approx175$, where $\Gamma\equiv Z^2 e^2/(a k_B T)=22.5Z^{5/3} x
/T_6$ is  the Coulomb coupling parameter, $T$ is temperature, $T_6\equiv
T/10^6\mbox{\,K}$, and $k_B$ is the Boltzmann constant \cite{PC00}. This
implies $T \lesssim 1.3\times10^5 Z^{5/3} x$~K, which leads to $T
\lesssim (3\times10^7)\,x$~K for an envelope of a neutron star composed
of iron or nickel or $T \lesssim (2.5\times10^6)\,x$~K for a carbon
plasma in a white dwarf or in an accreted crust of a neutron star.
Taking the electron polarization into account, one finds that the
crystallization may occur at somewhat smaller $\Gamma$ (larger $T$)
\cite{PC13}. Hence the Coulomb crystals can exist in the degenerate
stars not only at $x\gtrsim1$, but also in the nonrelativistic ($x<1$)
parts of envelopes of sufficiently old and cold neutron  stars (see,
e.g., Fig.~2.2. in Ref.~\cite{HPY07}) or white dwarfs (e.g., Fig.~4 in
Ref.~\cite{DAntonaMazzitelli}), where one can anticipate formation of
the fcc and hcp structures. 

However, the smallness of energy differences between these structures
makes it possible that thermal fluctuations destroy the long-range order
at the temperature of crystallization and thus make the structure of
some domains out of the true ground state. Therefore, the neutron star
crust and white dwarf matter at such low densities and temperatures may
consist of polycrystalline mixtures of different types of lattices or
even amorphous solid. Beside the influence on the elastic and breaking
properties of the crust, the presence of polycrystalline or amorphous
structures can suppress the electrical and thermal conductivities,
causing observable consequences for the magnetic and thermal evolution
of a neutron star. This possibility deserves a further study.

\begin{acknowledgments}

This research was supported by The Ministry of Science and Higher
Education of the Russian Federation (Agreement with Joint Institute for
High Temperatures RAS No.\,075-15-2020-785).

\end{acknowledgments}

\end{document}